# Characterisation of neutron field in large graphite insertion in special core of the LR-0 Reactor

Tomáš Peltan[*], Eva Vilímová[*], Tomáš Czakoj[†], Zdeněk Matěj, Filip Mravec, František Cvachovec, Jan Šimon[†], Vlastimil Juříček[†], Michal Košťál[†], Martin Schulc[†]

[*]*University of West Bohemia, Univerzitni 8, Pilsen, 301 00 Czech Republic*
[†]*Research Centre Rez, Hlavní 130, Řež, 250 68 Czech Republic*
*Masaryk University, Botanická 68a, Brno 602 00, Czech Republic*

**Abstract**

*Graphite is important reactor material used as a neutron moderator in various reactor designs. It is not only structural material used in previous reactors design but even material to which attention is focused due to the very high-temperature reactor design, which uses graphite as moderator and reflector. Due to the well-characterised cross-section of graphite, it is also a good candidate for neutron shaping experiments. In this kind of experiment, the neutron spectra are modified to the unconventional shape of neutron spectra which can be used to validate cross-section. This paper deals with validation of neutron field in graphite block surrounded by driver core arranged in large special core in LR-0 reactor. The experimentally determined shape of neutron spectra and spatial distribution of neutron flux is compared with the calculation. It is worth noting that the agreement for flux distribution is satisfactory in regions below the moderator level.*

## 1 Introduction

In recent years, reactor graphite has become a more popular material. It can be used as a neutron reflector or a moderator in nuclear reactors. Graphite is a crucial material for a wide range of common nuclear reactor types such as Gas Cooled Reactors (GCR) and Light Water Graphite Reactors (LWGR) and even for newly developed reactors such as Molten Salt Reactors (MSR) or other currently very popular Small Modular Reactors (SMR). As the world moves towards decarbonisation and the demand for SMRs and graphite reactors increases, developing new reactor concepts should be supported by adequate experimental research to ensure the credibility and quality of these designs. Although the behaviour and neutron properties of the graphite are well known, there are still some discrepancies in graphite microscopic cross-sections (Brown et al., 2018).

Moreover, there is a lack of well-characterised neutron fields in graphite that can be used for experimental purposes. Therefore, this article deals with a preliminary characterisation of the graphite neutron field in a special large graphite insertion in the LR-0 experimental reactor. With the unique graphite insertion, the LR-0 reactor core could serve as a new reference neutron field for various experiments requiring a precise thermal and epithermal neutron spectrum. Once the final neutron graphite field in LR-0 is well characterised, it can be used for various important experiments such as cross-section verification, neutron detector response evaluation, and other graphite-related studies.

This paper presents the results of a neutron distribution mapping performed using a set of activation foils, which were selected according to their suitable microscopic cross-section for thermal and epithermal neutron field reconstruction to characterise the neutron field of graphite prism. Then the paper analyses a fast neutron spectrum measured by the stilbene detector above 1 MeV with the stilbene detector and compares the experimental data from measurement in the LR-0 reactor with calculations performed in Monte Carlo codes MCNP, Serpent, and SCALE.

## 1.1 Experimental reactor LR-0

An irradiation experiment was performed in a specially developed reactor core assembled in the LR-0 reactor. The LR-0 is a zero-power, light water moderated, pool-type reactor operated at atmospheric pressure and room temperature in the Research Centre Rez. The reactor uses mainly VVER-1000 fuel with a fission column height of 125 cm. The main purpose of the reactor is the research and development of VVER-1000 and VVER-440-based reactor cores and benchmark experiments. The characteristics of the reactor enable a wide range of experiments and different core arrangements. The criticality of each core arrangement is reached by changing the moderator level or control rod position. The continuous nominal power of LR-0 is 1 kW with thermal neutron flux $\approx 10^{13}$ m$^{-2}$ s$^{-1}$. Figure 1 shows the schematic view of the reactor.

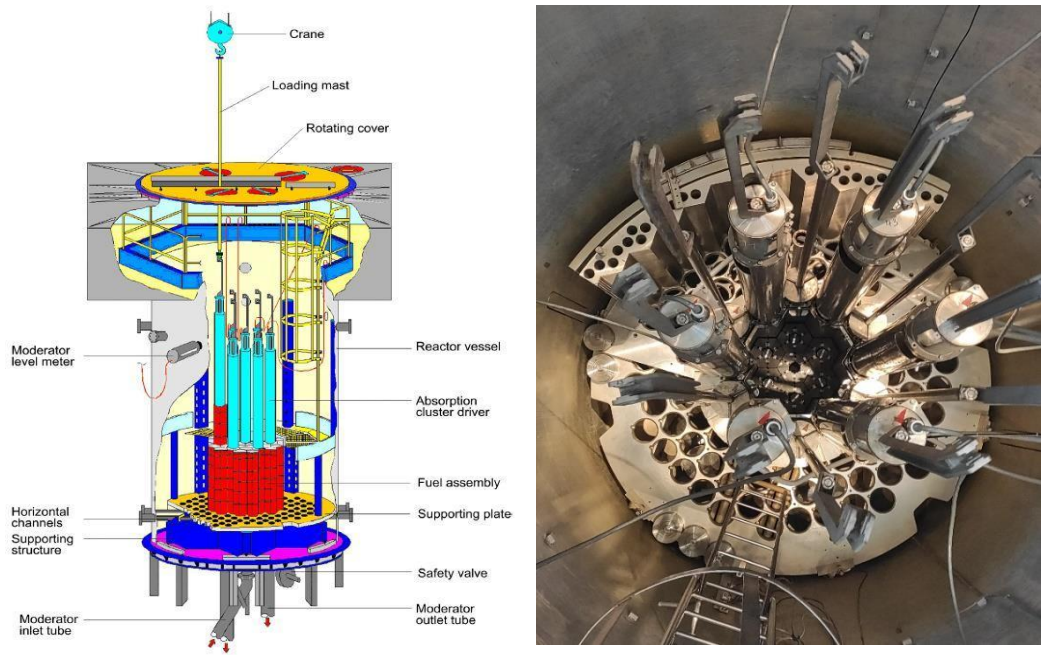

*Figure 1: LR-0 reactor general view (left side), photo of real reactor arrangement (right side)*

## 2 Experimental methods

### 2.1 Experiment core arrangement

The LR-0 reactor core, in the presented irradiation experiments, consists of twelve fuel assemblies with a nominal enrichment of 3.59 % and 3.60 % of $^{235}$U. The fuel pins have a pin pitch equal to 1.275 cm, and fuel assemblies have a pitch formed of 23.6 cm. The experimentally determined critical level of the moderator reaches 39.133 ± 0.005 cm in this core arrangement. A neutron flux measurement is ensured by nine dry aluminium channels surrounding the core installed 4 cm away from the core in a light water reflector.

Twelve fuel assemblies surround a specially developed new dry experimental module, corresponding to seven fuel assemblies in radial size and shape. The dry experimental module is made of pure aluminium walls and assembled with stainless steel bottom weights and a support socket located in the lower part of the whole module. The dry module support socket fits to fuel construction supports of the reactor. Compared with previous separate experimental modules presented in previous papers (Sobaleu et al., 2021; Czakoj et al., 2018), the module used for this article has an important advantage. Its complex geometry allows the placement of the graphite blocks into the dry module in tight geometry without significant air gaps, which keeps the module dry during all phases of the experiment. This feature, together with the addition of central graphite cylinder plugs, ensures a very "clean" neutron field dependent only on the embedded material because there is no additional excess structural material or water gaps between the separate experimental modules. The visualisation of the reactor core, with developed large graphite insertion in the dry experimental module, can be seen in Figure 2.

The experimental dry module is filled with seven identical graphite blocks, and each block consists of six smaller trapezoidal parts with a central cylinder in tight geometry. These seven blocks were installed on a special aluminium base for easier handling. The height of each graphite block is 60 cm with a hexagonal key (flat-to-flat) dimension of 21.65 cm. The graphite insertion used in these experiments has a density of 1.72 ± 0.02 g/cm$^3$ and a concentration of impurities below 0.2 ppm of boron equivalent and thus meets the nuclear graphite limits (Bolewski et al., 2005). The central position of the graphite cylinder has been replaced by a unique activation holder made of pure aluminium, which is used to insert the activation foils fixed in the exact place of the holder.

## 2.2 Activation foils arrangement

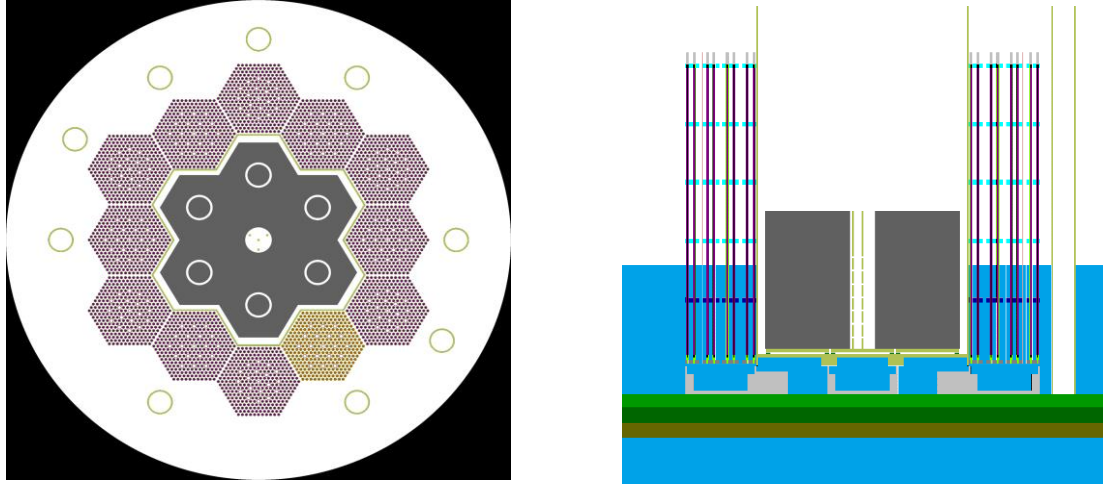

*Figure 2: Schematic view of the reactor arrangement - floor plan (left) and side view (right).*

A suitable combination of Au, Cu, Fe, Mn, and Ta activation foils then provides the required accurate mapping of the thermal neutron distribution. The golden activation foils had two modifications, one being 100 % Au and the other aluminium alloy mixed with 1 % Au. The advantage of 1 % Au activation foil is the almost negligible effect of a resonance self-shielding effect. Therefore, the self-shielding correction factor is very low. In an effort to map epithermal neutrons, several foils were encapsulated in a cadmium filter that cuts off almost all thermal neutrons. The method with cadmium filter is widely used and verified (Košťál et al., 2021). The actual arrangement and precise position of the activation foil fixed to the holder can be seen in Figure 3, Figure 4, and Table 1. Activation foils had different sizes and shapes. The 100 % Au and 1 % Au activation foils had square dimensions of 3×3×0.05 mm. Iron and copper foils were square as well, with the dimensions 10×10×3 mm for iron and 3×3×0.1 mm for copper activation foil. The manganese was normalised to a small sphere with a radius of 0.0761 mm, and tantalum was a cylindrical shape with a radius equal to 1.8 mm and height equal to 0.01 mm.

The unique activation foil holder consists of four pure aluminium rods with a diameter of 0.8 cm assembled in defined positions. A distance between each activation material was set to 5 cm in each rod to prevent any neutron field disruption and possible interference between the materials. Irradiation of the activation foils was performed in two independent experiments. In the first experiment, the irradiation lasted for two days, with the first part of the irradiation taking 9 hours and the second part taking 7 hours. Both experiments were operated at the same power of approximately 5 W. The second experiment was performed with such selected materials, whose absorption reactions are considered as a reference for thermal neutron mapping, to verify the thermal neutron flux distribution reconstruction. The second set of activation films was irradiated for 5.5 hours at approximately 5 W.

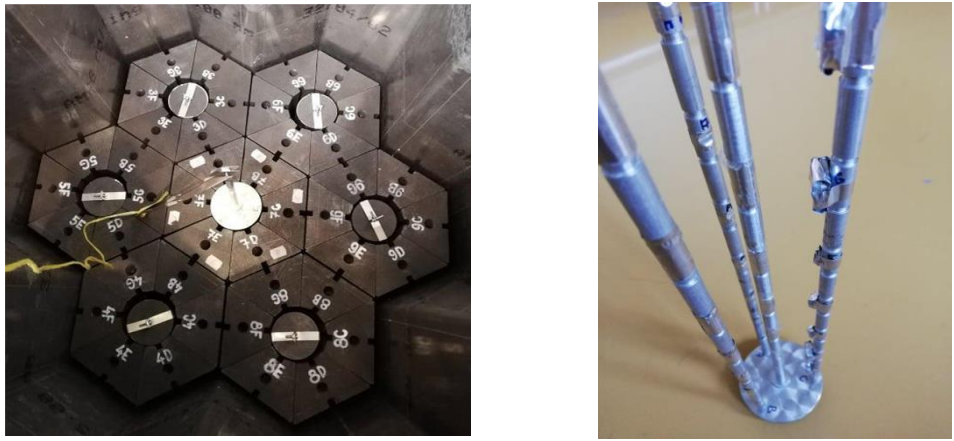

*Figure 3. Photo of real arrangement in the reactor core (left side) and the activation foil*

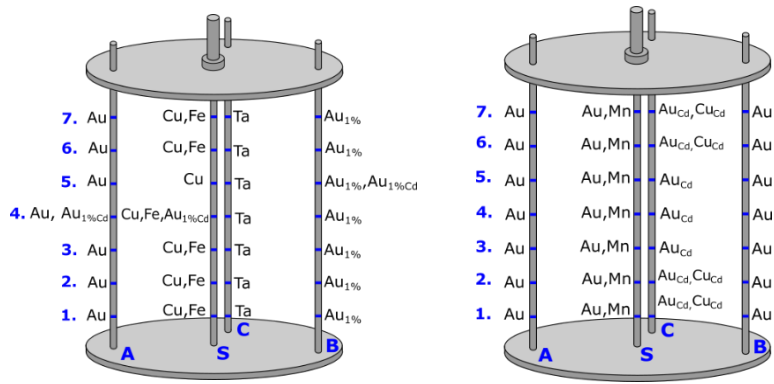

*Figure 4. Schematic of the aluminium activation foil holder – the layout in the first experiment (left) and the arrangement in the second experiment (right)*

*Table 1. Activation foil positioning in the activation foil holder during both experiments, (Cd) stands for cadmium filter*

| Axial position | Holder rod S | Holder rod A | Holder rod B | Holder rod C |
|---|---|---|---|---|
| 1 | Cu, Au, Fe, Mn | Au | $_{1\%}$Au, Au | Au $_{(Cd)}$, Cu $_{(Cd)}$, Ta |
| 2 | Cu, Au, Fe, Mn | Au | $_{1\%}$Au, Au | Au $_{(Cd)}$, Cu $_{(Cd)}$, Ta |
| 3 | Cu, Au, Fe, Mn | Au | $_{1\%}$Au, Au | Au $_{(Cd)}$, Ta |
| 4 | Cu, Au, Fe, $_{1\%}$Au $_{(Cd)}$, Mn | Au, $_{1\%}$Au $_{(Cd)}$ | $_{1\%}$Au, Au | Au $_{(Cd)}$, Ta |
| 5 | Cu, Au, Mn | Au | $_{1\%}$ Au, $_{1\%}$Au $_{(Cd)}$, Au | Au $_{(Cd)}$, Ta |
| 6 | Cu, Au, Fe, Mn | Au | $_{1\%}$Au, Au | Au $_{(Cd)}$, Cu $_{(Cd)}$, Ta |
| 7 | Cu, Au, Fe, Mn | Au | $_{1\%}$Au, Au | Au $_{(Cd)}$, Cu $_{(Cd)}$, Ta |

## 2.3 Determination of Reaction Rate

After the irradiation in the reactor core, the activity of irradiated activation foils was measured using a well-defined HPGe detector (see Figure 9 in (Košťál et al., 2018)). The HPGe detector measures the Net Peak Areas (NPA) of irradiated activation foils in a vertical configuration (Ortec GEM35), and the HPGe efficiency calibration was calculated using the detector model in MCNP6.2 code (Werner, 2018). The model was prepared using experimentally determined HPGe dimensions obtained by the detector radiogram (Dryak et al., 2006) and precisely measured dead layer (Boson et al., 2008). The calculations with a fixed source of neutron spectrum were used to determine the resonance self-shielding factors. These neutron spectra were obtained for each position in the activation holder, and detailed spectra were obtained by the criticality calculation model of the reactor core.

The reaction rate of radioisotopes originating during non-constant irradiation is derived using Equations (1) and (2), while the capture reactions of the produced radioisotopes are neglected.

$$\frac{A(\underline{P})}{A_{Sat}(\underline{P})} = \sum P_{rel}^i \times \left(1 - e^{-\lambda.T_{ir}^i}\right) \times e^{-\lambda.T_{end}^i} \quad (1)$$

$$q(\underline{P}) = \left(\frac{A(\underline{P})}{A_{Sat}(\underline{P})}\right)^{-1} \times NPA(T_M) \times \frac{\lambda}{\varepsilon \times \eta \times N} \times \frac{\frac{t_{real}}{t_{live}}}{(1-e^{-\lambda.T_m})} \times \frac{1}{e^{-\lambda.\Delta T}} \times \frac{1}{k_{CSEF}} \times k_{SSEF} \quad (2)$$

where:

$\frac{A(\underline{P})}{A_{Sat}(\underline{P})}$ is relative portion of saturated activity induced during irradiation experiment,

$P_{rel}^i$ is relative power on the i-th day of irradiation, $P_{rel}^i = \frac{P^i}{P}$,

$q(\underline{P})$ is reaction rate of activation foil during power density $\underline{P}$,

$T_{ir}^i$ is irradiation time on i-th day of irradiation,

$T_{end}^i$ is time from the end of i-th day of irradiation to end of all irradiations,

$\lambda$ is decay constant of corresponding material,

$T_m$ is the time of activation foil measurement by HPGe,

$\Delta T$ is the time between the end of irradiation and the start of HPGe measurement,

$NPA(T_m)$ is the measured number of counts,

$\varepsilon$ is gamma branching ratio of activation material – depending on the material,

$\eta$ is detector efficiency - the result of MCNP calculation,

$N$ is number of target isotope nuclei in activation foil,

$t_{real}$ is the real-time of counting system of the HPGe (= $T_m$),

$t_{live}$ is the live time of the counting system of the HPGe (< $t_{real}$),

$k_{CSEF}$ is the coincidence summing effect correction,

$k_{SSEF}$ is the resonance self-shielding effect correction determined by MCNP.

The previous research (Košťál et al., 2018) demonstrates the applicability of such an approach. More details about the measurement methodology and whole setup can be found in (Košťál et al., 2021). The measurement parameters of the activation foils and analysed reactions of the nuclides are shown in Table 2.

*Table 2. Reactions and parameters of analysed activation samples*

| Measured Nuclide | Concentration | Peak Energy (keV) | HPGe Efficiency | $k_{CSEF}$ | $k_{SSEF}$ | $k_{SSEF}$ in Cd |
|---|---|---|---|---|---|---|
| $^{197}$Au(n,γ)$^{198}$Au | 100% | 411.8 | 8.22E-02 | 0.998 | 1.778 | 3.539 |
| $^{197}$Au(n,γ)$^{198}$Au | 1% | 411.8 | 8.22E-02 | 0.998 | 1.009 | 1.022 |
| $^{63}$Cu(n,γ)$^{64}$Cu | 100% | 511.0 | 6.71E-02 | 1.000 | 1.054 | 1.427 |
| $^{58}$Fe(n,γ)$^{59}$Fe | 98% | 1099.2 | 3.36E-02 | 0.988 | 1.062 | 1.067 |
| $^{55}$Mn(n,γ)$^{56}$Mn | 100% | 846.7 / 1810.7 | 2.39E-03 | 0.996 / 0.990 | 1.081 | - |
| $^{181}$Ta(n,γ)$^{182}$Ta | 100% | 1121.0 / 1221.4 | 3.56E-02 | 0.867 / 0.940 | 1.911 | - |

## 2.4 Stilbene detector measurement

Fluxes in the energy range above 1 MeV are measured with a two-parameter multichannel analyser, nuclear electronics for detector signal processing (preamplifier, high-voltage supply, analyser) stilbene scintillator of cylindrical geometry with dimensions ø 10 mm × 10 mm and photomultiplier RCA 8575. Because the scintillations in organic scintillation detectors are caused not only by neutron interactions but also by gamma-ray interactions, the separation between neutron and gamma pulses is realised by means of pulse shape discrimination (PSD) of the measured response. Pulse Shape Discrimination parameter (D) is derived by an integration algorithm, which principle lies in the comparison of area limited by part of a trailing edge of the measured response (*Q1*) with area limited by the whole response (*Q2*). The areas *Q1* and *Q2*, as integrals over time, are expressed in Equation (3), and their illustration is shown in Figure 5.

The time offset $t_2$ is set for the optimal discrimination properties (namely, the most significant possible difference in the discrimination parameter for neutrons and gammas) to about 1/10 to 1/3 of the trailing edge. It varies for each scintillation material (stilbene, p-terfenyl). In this way, it is possible to eliminate the classification mistakes caused by the dependency of the response shape on its amplitude.

Charge $Q_1$ is determined by an area limited by the response course within a time interval ($t_2$, $t_3$). The charge $Q_2$ is determined by an area limited by the response course within firmly defined times $t_0$ and $t_3$. Times $t_0$ and $t_3$ depend on the parameters of the measuring apparatus, and time $t_3$ is defined as the end of the response.

Using PSD, energy-dependent recoil proton responses $S(E_p)$ are evaluated. The neutron fluxes are then evaluated by deconvolution according to Equation (4). The response matrix of the crystal $K(E_N, E_P)$ was determined employing Monte Carlo code NEU-7. The methodology was tested in various fields, especially fission reactor fields, $^{252}$Cf(s.f), special fields, and also DT generator fields (Košťál et al., 2018, Schulc et al., 2022, Košťál et al., 2017, Czakoj et al., 2022).

$$Q_1 = \int_{t_2}^{t_3} i(t)dt, \quad Q_2 = \int_{t_0}^{t_3} i(t)dt, \quad D = \frac{Q_1}{Q_2} \qquad (3)$$

$$S(E_p) = \int K(E_N, E_P)\phi(E_N)dE_N \qquad (4)$$

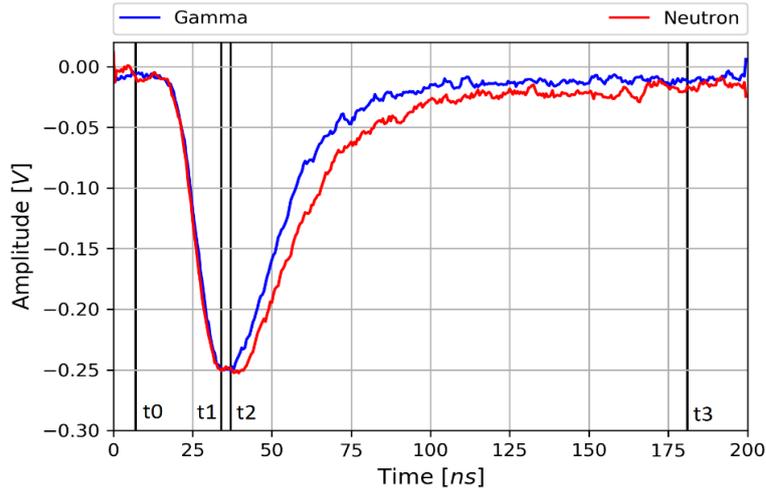

*Figure 5.: Comparison of real neutron and gamma pulses from Stilbene scintillation detector with marked examples of a separation boundary for integration algorithm*

## 3 Calculations

### 3.1 MCNP and Serpent calculations

The neutron transport and neutron spectra simulations were carried out using the MCNP 6.2, Serpent 2.1.31 (Leppänen et al., 2015), and SCALE 6.2.3 (Wieselquist et al., 2020) Monte Carlo codes with various nuclear data libraries. An accurate model has been developed for all the cases studied, the only simplification being the neglect of a small air gap in each graphite block. The simplification was done by homogenising the air cavity and graphite blocks by changing the density of the whole graphite module to ensure the same amount of graphite in each module, see the comparison in Figure 2 (model) and Figure 3 (photo).

Neutron spectra were calculated at specific positions of the activation holder in two computational steps. First, the raw neutron spectra at the holder location were calculated, and then the cadmium-filtered spectra were determined. These calculations were carried out only in MCNP6.2 code with ENDF/B-VII.1 nuclear data library and corresponding TSL (Thermal Scattering Library). Both calculations were in standard critical calculation mode with a fixed moderator level determined from critical experiments. Analyses of activation detectors with or without cadmium shaping filters were performed with 40,000 neutrons per cycle in 585,000 active cycles with 50 inactive cycles. The neutron spectrum was calculated for all 28 positions in the aluminium activation foil holder (four holder rods with seven axial irradiation positions each). The statistical uncertainty of the calculated neutron spectrum, reached in the central position *S3* of the holder, is below 2% in each energy group in the energy range from $1\times10^{-8}$ MeV to 3 MeV. In higher and lower energy regions, the statistical uncertainty in each energy group slightly rises. Obtained neutron spectrum serves as a defined neutron spectrum for reaction rate calculation in previously described activation materials.

Reaction rates of all activation foils were calculated separately with a fixed neutron spectrum source and the exact shape of the activation detector in the separate MCNP calculation. In each calculation step, $1\times10^9$ source neutrons were simulated with an energy spectrum corresponding to the precise position in the activation foil holder. The simulated geometry was a sphere in which the previously calculated neutron source spectrum was simulated. The source sphere was 8 cm in diameter, and the actual specific activation foil with the corresponding shape and dimensions was placed in the centre of the sphere. The statistical uncertainty in the calculated reaction rate was between 0.3% and 0.7%, depending on the activation material. The geometry of all activation detectors was calculated with and without the activated material. Finally, the correction factors and positions of all materials were determined based on the obtained reaction rates for the specific material and cavity geometry. The self-shielding factor was defined as the share of two reaction rates: the reaction rate obtained from the calculation without activation material divided by the reaction rate obtained from the calculation with activating material. These self-shielding correction factors were later used to modify experimentally obtained data and link calculation and experiment. After that, reaction rates of all materials used for neutron mapping in the activation foil holder and comparison with the experiment were manually calculated using the scalar multiplication of the calculated neutron spectrum in a defined position and microscopic cross-section obtained from ENDF/B-VIII.0 nuclear data library (Brown et al., 2018). This technique determines the reaction rate per atom in each activation foil.

MCNP and Serpent were also used to calculate a wide range of the neutron spectrum. The ENDF/B-VIII.0 nuclear data library was used with a corresponding TSL matrix for this calculation. In the case of graphite, the correct TSL with 30% graphite porosity was used, which is consistent with previous research on graphite calculations and experiments performed on the LR-0 reactor (Košťál et al., 2016). The selected porosity of graphite best corresponds to the real density and structure of the graphite used for these experiments. The MCNP calculation performed 200,000 neutrons per cycle in 20,000 active and 50 inactive cycles. The uncertainty of the calculated neutron spectrum is below 0.8% in energy ranges from $1\times10^{-8}$ MeV to 3 MeV. The same task was calculated using the Serpent code. For this calculation, 200,000 neutrons per cycle were simulated in 80,000 active cycles with 50 inactive cycles. The nuclear data library and other parameters such as temperature were the same as in the previous MCNP calculations. To reach the same uncertainty as MCNP, a higher number of simulated generations was chosen in Serpent simulations. The need for a larger number of computational cycles was probably due to the different methodology of modelling the whole system in the Serpent code compared to MCNP. The simulated detection volume was modelled at the centre of the graphite insertion height, in the axial position corresponding to the measuring point of the Stilbene detector.

The behaviour of the neutron flux and the shape of the neutron spectrum have been described in previous sections of this paper by calculating the activation detector and stilbene measurements. The pin power distribution in the selected fuel assembly was calculated to understand better how the presence of graphite affects the power distribution throughout the reactor core. The pin power distribution was calculated across the one fuel assembly in ninety axial layers for eight defined fuel pins (see Figure 7) using tally F7 calculation mode. The F7 tally, as one of the standardised tallies in MCNP, calculates fission energy deposition averaged over all cells in units of MeV/g. Obtained results show averaged fission power in defined cells which directly corresponds to the neutron flux in the observed cell. The exact axial division can be found in the results in Figure 8. This calculation was performed only in MCNP as a criticality calculation with the same nuclear data library as in the previous calculations performed and with 100,000 neutrons in 50,000 active cycles, of which 50 cycles were inactive as in the previous cases. The effect of inserted graphite is described in detail in the results section. The obtained results of fuel assembly power shaping can be compared with core power distribution in the reference neutron field (Košťál, 2020).

### 3.1.1 Sensitivity analyses in SCALE

Sensitivity was calculated using the SCALE6.2.3/TSUNAMI-3D code (Rearden & Jessee, 2017). Simulations were done by applying the Iterated Fission Probability method (Perfetti et al., 2016) and the ENDF/B-VII.1 continuous energy nuclear data library. Simulations employed 15,000 generations with 50,000 neutrons per generation, and 100 generations were skipped. The validity of the results was confirmed by the direct perturbation approach for the most sensitive isotopes.

## 4 Results

## 4.1 Determination of critical height of moderator level

The critical heights of moderator level ($H_{cr}$) measurements are summarised in Table 3. At this level, the reactor was critical, i.e., $k_{eff}$ = 1.00000. The effect of the graphite insertion can be seen compared to cases without graphite. All the measurements were performed three times to achieve higher accuracy of the results.

*Table 3: Experimental and calculated data of critical height of the moderator level $H_{cr}$ – ENDF/B-VII.1 nuclear data library*

| Modification of central module | Experimental $H_{cr}$ [mm] | MCNP calc. $k_{eff}$ [-] | Serpent calc. $k_{eff}$ [-] | SCALE calc. $k_{eff}$ [-] |
|---|---|---|---|---|
| With graphite insertion | 391.33 ± 0.05 | 0.99543 ± 0.00006 | 0.99635 ± 0.00006 | 0.99569 ± 0.00005 |
| Empty experimental module | 564.23 ± 0.09 | 1.00140 ± 0.00005 | 1.00189 ± 0.00006 | 0.99863 ± 0.0005 |

The difference in moderator level between the case without graphite insertion and the case with graphite insertion is significant, indicating that graphite is a good moderator and reflector in the reactor core. One can notice that the case without graphite insertion agrees better with the experiment than the case with graphite insertion, which may be caused by disagreement in the graphite microscopic cross-section or in the TSL data library.

*Table 4: Results from MCNP calculations from various combinations of nuclear data libraries*

| Fuel and another material library | Graphite library | TSL library | $k_{eff}$ [-] ± 0.00006 |
|---|---|---|---|
| ENDF/B-VII.1 | ENDF/B-VII.1 | / | 0.99482 |
| ENDF/B-VII.1 | ENDF/B-VII.1 | ENDF/B-VII.1 | 0.99543 |
| ENDF/B-VII.1 | ENDF/B-VII.1 | ENDF/B-VIII.0 crystalline modification | 0.99770 |
| ENDF/B-VII.1 | ENDF/B-VII.1 | ENDF/B-VIII.0 10% porosity | 0.99787 |
| ENDF/B-VII.1 | ENDF/B-VII.1 | ENDF/B-VIII.0 30% porosity | 0.99785 |
| ENDF/B-VII.1 | JEFF 3.3. | JEFF 3.3 | 0.99539 |
| ENDF/B-VII.1 | ENDF/B-VIII.0 | / | 0.99487 |
| ENDF/B-VII.1 | ENDF/B-VIII.0 | ENDF/B-VII.1 | 0.99542 |
| ENDF/B-VII.1 | ENDF/B-VIII.0 | ENDF/B-VIII.0 crystalline modification | 0.99778 |
| ENDF/B-VII.1 | ENDF/B-VIII.0 | ENDF/B-VIII.0 10% porosity | 0.99782 |
| ENDF/B-VII.1 | ENDF/B-VIII.0 | ENDF/B-VIII.0 30% porosity | 0.99786 |
| ENDF/B-VII.1 | ENDF/B-VIII.0 | JEFF 3.3 | 0.99538 |
| ENDF/B-VIII.0 | ENDF/B-VIII.0 | ENDF/B-VIII.0 30% porosity | 0.99383 |

Various combinations of nuclear data libraries and different TSL, only for graphite, can be found in Table 4. For all cases, the construction and other core materials were modelled in the same library ENDF/B-VII.1. only the Ographite cross-section and TSL matrix were changed. Table 4 shows that when TSL in ENDF/B-VII.1 is used, the difference is only approximately 60 pcm compared to the case without the TSL matrix of graphite. On the other hand, a very strong influence of all TSL modifications from the ENDF/B-VIII.0 nuclear data library is observed, with more than 240 pcm difference. Compared to this phenomenon, the JEFF 3.3. TSL matrix for graphite achieved a very similar result to the ENDF/B-VII.1, indicating that the evaluation of ENDF/VIII.0 is not entirely correct due to high overestimations of achieved results in comparison with other cases.

## 4.2 Reaction rate measurement

Based on the previously mentioned methodology, the activity of the activation detectors was measured on the HPGe detector and then calculated based on Equations (1) and (2). Estimated reaction rates based on measurement were normalised using a scaling factor. The scaling factor is a calculated constant representing the neutron emission in the reactor core. For calculating the scaling factor from the experiments, the average value based on Ta and $_{1\%}$Au activation foils were used (positions 2, 3, and 4 in the activation foil holder). The scaling factor is calculated from the experimentally obtained reaction rate multiplied by the self-shielding correction factor and divided by the calculated reaction rate from the MCNP code calculation. The scaling factor for both experiments has been determined to be $4.245 \times 10^{11}$ with 0.86% statistical uncertainty. The average neutron emission per one fission obtained from MCNP calculation was 2.447 neutrons, with fission energy released of 180.9034 MeV. From these values, the irradiation power during the experiments mentioned above (see section 2.2) was calculated to be approximately 5 W. Table 5 shows experimentally determined reaction rates of the activation foils in the defined position on the activation holder. The statistical uncertainty of measurement has been quantified with a combination of measured geometry uncertainty and uncertainty of HPGe detector below 0.79 % for all observed activation detectors.

In order to compare experimental data and MCNP calculations of reaction rates, the C/E-1 (Calculation/Experiment -1) comparison was realised (see Table 6). It is shown an excellent agreement between measurement and calculation in positions 3, 4, and 5, which is caused by their position in the centre of the activation holder (Figure 4) in the centre of the graphite insertion in the axial direction. In contrast, positions 1, 2, and 6, 7 are underestimated for almost all detectors. This phenomenon may be caused by their axial position on the holder, which is, in the case of positions 1 and 2, at the bottom part of the graphite insertion module near the lower construction parts of the reactor core. The module has stainless steel weight and a relatively large amount of water, which seems to affect the shape of the neutron spectrum. At the same time, other boundary phenomena such as the end of the fuel column in this area may play a non-negligible role in neutron spectra shaping.

On the other hand, position 6 may be affected by the moderator-air interface, and position 7 is located above the moderator level. Position 7 shows the highest inconsistency of all the activation holder positions examined. As the difference between the moderator level and the position of the activation foil increases, the underestimation rises, corresponding to the neutron flux distribution previously observed (Košťál et al., 2016).

The neutron flux profile in graphite block is illustrated by a plot of the Au reaction rates along the height of the special activation holder in Figure 6. It is demonstrated relatively high uniformity of the neutron field in the radial direction in all irradiation positions, especially around the centre of the activation holder in positions 3, 4, and 5, indicating the high homogeneity of the neutron field in the radial direction.

*Table 5: Experimentally determined reaction rates of the activation foil per one atom in the exact position on the unique aluminium activation foil holder. The unit is [1/s]*

| Material | Au | Cu | Fe | Mn | Au $_{100\%}$ | Au $_{100\%}$ | Au $_{1\%}$ | Ta | Au $_{100\%}$ (Cd) | Cu (Cd) | Au $_{1\%}$ (Cd) | Au $_{1\%}$ (Cd) | Au $_{1\%}$ (Cd) |
|---|---|---|---|---|---|---|---|---|---|---|---|---|---|
| Position | Rod S | Rod S | Rod S | Rod S | Rod A | Rod B | Rod B | Rod C | Rod C | Rod C | Rod S | Rod A | Rod B |
| 1 | 7.650E-27 | 1.872E-28 | 5.406E-29 | 5.311E-28 | 7.844E-27 | 7.597E-27 | 7.976E-27 | 2.703E-27 | 3.718E-27 | 1.573E-29 | - | - | - |
| 2 | 8.441E-27 | 2.040E-28 | 5.990E-29 | 5.733E-28 | 8.831E-27 | 8.437E-27 | 8.990E-27 | 2.976E-27 | 4.500E-27 | 1.886E-29 | - | - | - |
| 3 | 9.171E-27 | 2.047E-28 | 5.996E-29 | 6.122E-28 | 9.431E-27 | 9.169E-27 | 9.510E-27 | 3.196E-27 | 5.016E-27 | - | - | - | - |
| 4 | 9.489E-27 | 2.083E-28 | 5.866E-29 | 6.290E-28 | 9.554E-27 | 9.493E-27 | 9.715E-27 | 3.235E-27 | 5.856E-27 | - | 5.623E-27 | 5.791E-27 | - |
| 5 | 8.990E-27 | 2.005E-28 | - | 6.077E-28 | 9.271E-27 | 9.253E-27 | 9.793E-27 | 3.261E-27 | 5.954E-27 | - | - | - | 5.696E-27 |
| 6 | 8.896E-27 | 1.912E-28 | 5.167E-29 | 5.511E-28 | 8.912E-27 | 8.935E-27 | 8.904E-27 | 2.962E-27 | 5.113E-27 | 2.008E-29 | - | - | - |
| 7 | 7.877E-27 | 1.698E-28 | 4.664E-29 | 4.830E-28 | 8.061E-27 | 7.736E-27 | 8.242E-27 | 2.353E-27 | 4.598E-27 | 1.882E-29 | - | - | - |

*Table 6: Calculated C/E-1 for all reaction rates in all positions on the holder*

| Material | Au | Cu | Fe | Mn | Au | Au | Au 1% | Ta | Au (Cd) | Cu (Cd) | Au 1% (Cd) | Au 1% (Cd) | Au 1% (Cd) |
|---|---|---|---|---|---|---|---|---|---|---|---|---|---|
| Position | Rod S | Rod S | Rod S | Rod S | Rod A | Rod B | Rod B | Rod C | Rod C | Rod C | Rod S | Rod A | Rod B |
| 1 | -0.01% | -9.64% | -8.88% | -5.22% | -2.77% | 0.98% | -3.81% | -5.71% | 12.55% | -11.61% | - | - | - |
| 2 | 5.49% | -4.88% | -5.87% | -0.05% | 0.51% | 6.55% | -0.01% | 1.18% | 10.05% | -11.34% | - | - | - |
| 3 | 5.24% | 0.10% | -0.86% | -1.01% | 3.10% | 4.25% | 0.52% | -0.26% | 4.07% | - | - | - | - |
| 4 | 1.31% | -2.68% | 0.57% | -4.41% | 0.73% | 0.67% | -1.63% | 0.25% | -7.75% | - | -2.46% | -4.95% | - |
| 5 | 0.85% | -6.07% | - | -7.90% | -2.24% | -0.66% | -6.14% | -6.31% | -12.62% | - | - | - | -10.55% |
| 6 | -10.58% | -13.98% | -7.45% | -11.55% | -10.07% | -26.56% | -26.30% | -8.17% | -10.88% | -23.20% | - | - | - |
| 7 | -17.03% | -21.08% | -16.61% | -18.04% | -18.59% | -14.01% | -19.29% | -3.90% | -15.74% | -31.94% | - | - | - |

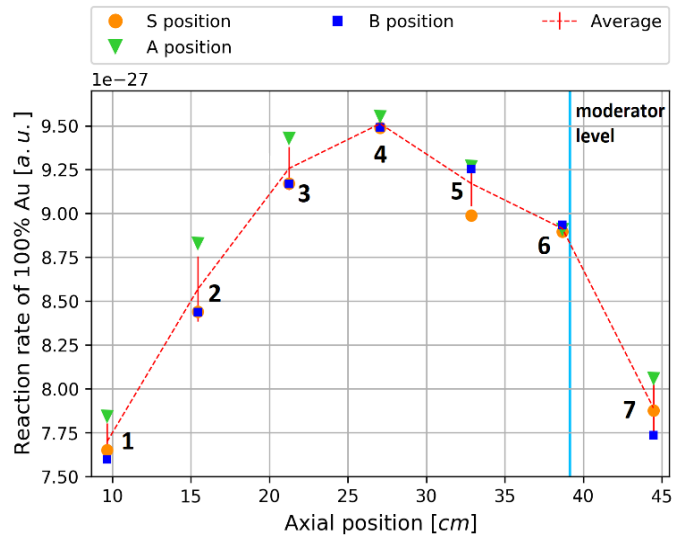

*Figure 6. Thermal neutron flux - radial profile in different axial positions of holder*

### 4.3 The effect of graphite on the neutron flux distribution

Experimental data and calculations have also demonstrated the effect of graphite on the neutron flux distribution in the axial direction of the reactor core. Figure 8 shows the axial profile of the neutron flux over the height of the reactor core. For illustration, the fission reaction rate in eight selected pins in the radial direction of the chosen fuel assembly was calculated, see Figure 7. Pin 1 is the closest to the experimental graphite module, and pin 8 stands on the other side of the fuel assembly at the furthest position. Additionally, the reaction rate distribution was calculated in the centre of the graphite insertion and compared to the experimental data obtained from the average reaction rate of Au activation foils.

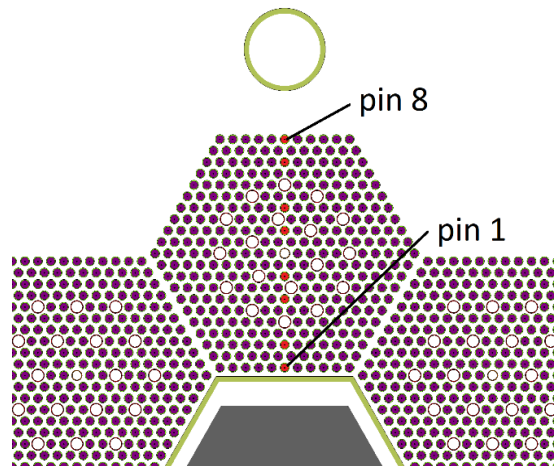

*Figure 7. Position of eight pins across fuel assembly for power distribution mapping*

The effect of the graphite presence in the centre of the reactor core can be seen in Figure 8. The influence is directly apparent in the neutron flux distribution of the fuel rods. The contribution of the neutron flux above the moderator level in the closest pin to graphite (pin 1) is higher than in pin 8 (pin 8), which is the farthest away from the graphite module. This effect, observed above the moderator level ($H_{cr}$ = 39.13 cm), is attributed to the graphite reflector because, below the moderator level, this effect is almost negligible. With increasing distance above the moderator level, the neutron flux decreases. Over the graphite blocks, however, the neutron flux decreases for both fuel pins with the same trend. The end of the graphite blocks corresponds to the height of 64.4 cm in Figure 8.

One can notice a considerable influence of the spacer grids located roughly in the centre of the reactor core from the axial course of the neutron flux in the fuel pins. Spacer grids strongly distort the shape of the neutron flux distribution in the fuel pins, and therefore the predicted maximum is distorted and shifted by this effect. Compared to neutron flux calculation in the graphite, an upward shift of the neutron flux maximum compared to the neutron flux distribution in the fuel can be observed, which is probably caused by the geometry of the graphite blocks and their position relative to the fuel column. In the centre of the graphite, the difference in the flux distribution between calculation results and measurement is minimal. As in previous measurement results, Figure 8 confirms good agreement at positions 3, 4, and 5 on the activation holder. Disagreement in positions 6 and 7 can be caused by neutron field non-uniformity and the angular distribution of the neutron field. There is almost no source of neutrons in the radial direction from graphite blocks. Positions 1 and 2 can be affected by the edge effects of steel and water located under the graphite insertion. Below the moderator level, the neutron flux profile in graphite is very close to a parabola. Above the moderator level, a linear decrease can be observed, which is consistent with previous assumptions.

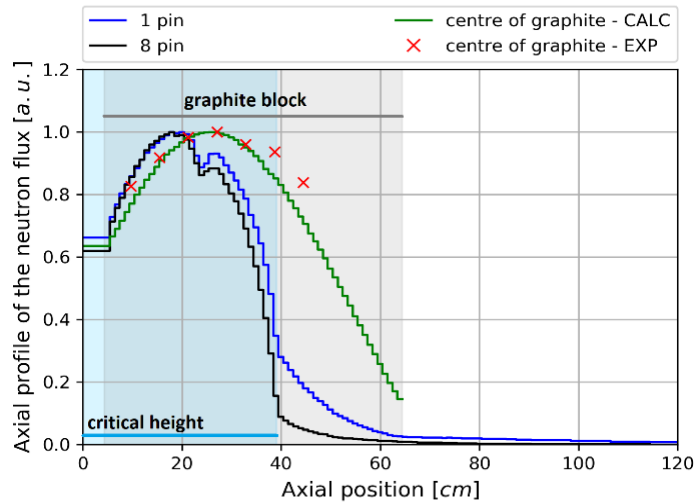

*Figure 8. Neutron flux distribution in the axial profile of the fuel and graphite insertion*

### 4.4 Fast neutron spectrum evaluation

The further stage was to evaluate the neutron spectrum in the fast energy region. Since the activation foils were selected with a focus on reactions in thermal and epithermal energy regions, a measurement with a stilbene scintillator was used to determine the fast neutron spectrum. The assumption determined by the calculation is a very well thermalised neutron spectrum in the proposed reactor core with graphite insertion, as shown in Figure 9. The total neutron spectrum was calculated using MCNP and Serpent in the centre of the graphite insertion and then compared with the neutron spectrum of the reference core without graphite (Košťál M. et al., 2020) determined in MCNP. Serpent and MCNP calculations are in excellent agreement, and obtained results were normalised to 1 by integrating over the entire neutron spectrum. In contrast to the reference LR-0 core, significant thermalisation occurs in the proposed reactor core with a thermal neutron ratio of almost 25% and a fast neutron ratio of less than 4% above 1 MeV. Moreover, the epithermal region is flat and nearly constant compared to the LR-0 reference core without the graphite central module, which can be further used to evaluate microscopic cross-sections in this energy region.

The calculated fast neutron spectrum of the reactor core was experimentally verified by measurement using the stilbene detector in the centre of the graphite insertion. For this purpose, calculated and measured data are normalised to 1 and integrated over the spectrum from 4 MeV to 10 MeV. The measurement was accomplished in the range from 1 MeV to 10 MeV, and Figure 10 reveals the measurement results and compares the results with calculations. Calculated results obtained from MCNP and Serpent calculations were modified by Gaussian broadening function using experimentally determined parameters for the stilbene detector. This method allows the comparison of experimental and calculated results. In the resulting calculated spectrum of fast neutrons, there are several sharp resonances due to resonances in the microscopic cross-section of graphite, namely between 2 MeV and 3 MeV. Considering the shape of the resonances, which are very sharp and narrow, they cannot be observed by the stilbene detector in detail, as this does not allow its resolution. The effect of these resonances can still be seen in Figure 9. It should be noted that the spectrum from Serpent is underestimated across the entire spectrum. On the other hand, Figure 9 suggests that Serpent is slightly overestimated in thermal energy.

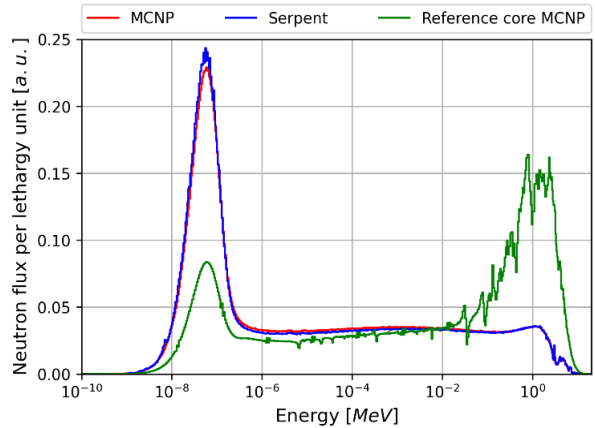

*Figure 9. Calculated neutron spectrum of the proposed reactor core with the graphite insertion compared with the reference LR-0 reactor core*

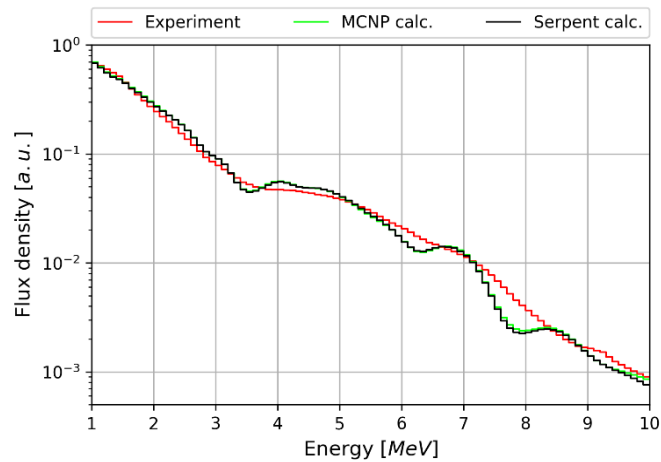

*Figure 10. Comparison of the experimentally measured fast neutron spectrum and spectrum calculated by MCNP and Serpent code compared with the reference LR-0 reactor core*

The calculated and measured results match very well from 1 MeV to 5.8 MeV. The deviation between the calculated and measured neutron spectrum increases from 6 MeV to 7 MeV, and around 8.5 MeV, the deviation is significant. Then, up to 10 MeV, the results are again in good agreement. To better understand the behaviour of the measured and calculated spectrum, Figure 11 compares the experimental data and calculation results in C/E-1. One can notice that, with a few exceptions, the region from 1 MeV to 5.8 MeV is within the +/- 15% uncertainty, and the discrepancy rises to -27% above the 5.8 MeV in specific troublesome energies. The most significant disagreement between the calculated and experimentally determined spectra may be observed around 7.8 MeV, where MCNP and Serpent calculations are underestimated by almost 50%, whereas the same tendency was observed in the previous experiments (Peltan T. et al., 2019). That massive discrepancy may be caused by certain evaluation inaccuracy in graphite microscopic cross-sections, mainly around 8 MeV. The criticality calculations of the reactor core with central graphite insertion, consistently underestimated throughout all reactor core calculations (Table 3), may also indicate this phenomenon. Based on these assumptions, a sensitivity analysis of microscopic cross-sections was created to illuminate this issue using TSUNAMI-3D.

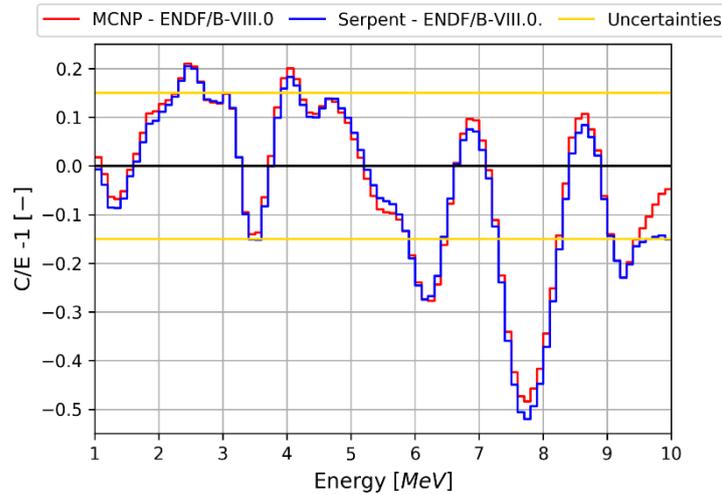

*Figure 11. The C/E-1 comparison of the fast neutron spectrum calculated by MCNP and Serpent and experimental data*

## 4.5 Sensitivity analyses by TSUNAMI

The results showed that graphite has one-tenth the sensitivity compared to the moderator, which is the most sensitive material in the core. It also showed that the fuel sensitivity is about 40% of the moderator sensitivity. The sensitivity shows how the $k_{eff}$ changes if the macroscopic cross-section of a given reaction (or material) is changed by 1%. The ten most significant reactions are tabulated in Table 7. It is not surprising that these are the reactions in the fuel and moderator.

Analysis of graphite reactions, tabulated in Table 8, has shown that elastic scattering is the dominant reaction. This reaction is the tenth most crucial reaction in the core. The second most important reaction in graphite, radiative capture, has a sensitivity lower than 5 % of elastic scattering sensitivity. Sensitivity profiles of both reactions and inelastic scattering are depicted in Figure 12.

*Table 7: The most sensitive reactions in the reactor core*

| Material | Nuclide | Reaction | Sensitivity [-] | Uncertainty |
|---|---|---|---|---|
| Fuel | $^{235}$U | nubar | 9.49E-1 | 0.00 % |
| Moderator | $^{1}$H | (n,n) | 3.46E-1 | 0.14 % |
| Fuel | $^{235}$U | (n,f) | 3.22E-1 | 0.02 % |
| Fuel | $^{238}$U | (n,γ) | -1.68E-1 | 0.01 % |
| Fuel | $^{235}$U | (n,γ) | -1.24E-1 | 0.01 % |
| Moderator | $^{1}$H | (n,γ) | -9.42E-2 | 0.02 % |
| Fuel | $^{238}$U | nubar | 5.06E-2 | 0.04 % |
| Moderator | $^{16}$O | (n,n) | 3.93E-2 | 0.36 % |
| Fuel | $^{238}$U | (n,f) | 3.42E-2 | 0.06 % |
| Graphite | Graphite | (n,n) | 3.12E-2 | 0.72 % |

*Table 8: Calculated sensitivities in the graphite*

| Reaction | Sensitivity [-] | Uncertainty |
|---|---|---|
| (n,n) | 3.12E-2 | 0.72% |
| (n,total) | 3.00E-2 | 0.75% |
| (n,γ) | -1.40E-3 | 0.07% |
| (n,n') | 2.38E-4 | 1.26% |
| (n,α) | -6.66E-5 | 0.56% |

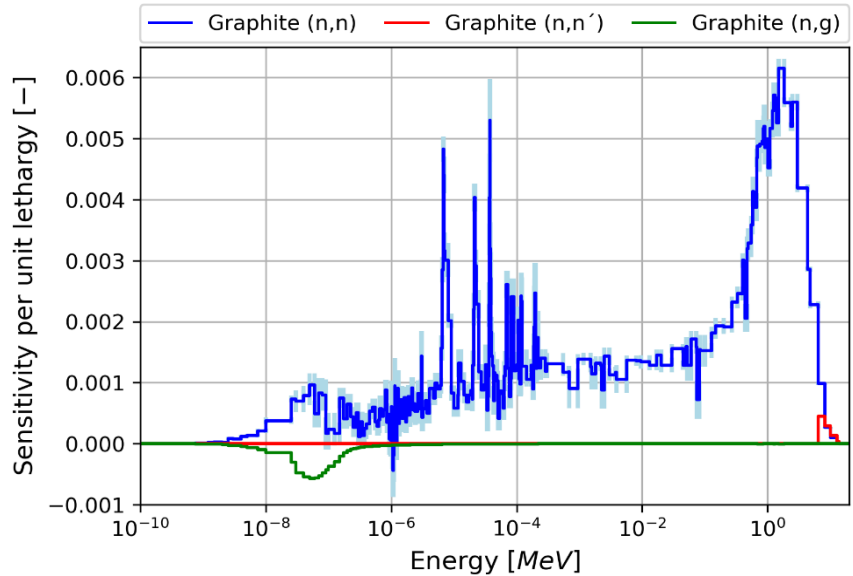

*Figure 12: Sensitivity per unit lethargy for the most important graphite reactions*

# 5 Conclusion

Sets of experiments in an LR-0 reactor with a large graphite insertion were performed to validate criticality, neutron flux distribution and neutron spectra in the large graphite insertion. The criticality was calculated in three independent codes, MCNP, Serpent, and SCALE. The experiment with a void central block was also realised for evaluating the graphite effect in the reactor core. It is worth noting that a good agreement was reached in the case with the void central block, which confirms a satisfactory description of the driver core. Discrepancies were observed when the central block was filled with graphite. The effect of the TSL matrix and graphite cross section on the result of the calculated criticality was analysed. The highest inconsistency compared to the other cases, which are very consistent, was achieved using TSL matrices from the ENDF/B-VIII.0 nuclear data library, which have a multiplicatively stronger influence than the other data libraries. That could imply possible discrepancies in graphite cross-section description. In this large block, approx. 60 cm thick, also inelastic scatter manifests itself on neutron transport, but in terms of the effect on criticality, the effect is almost negligible.

The problems in the graphite cross-section description might also imply the problems with the agreement between measured and simulated neutron spectra. In the region from 7 MeV to 9 MeV, the observed difference between measured and calculated spectra is as high as 50 %. It is worth noting that a similar manner of calculation to experiment agreement was observed in the measurement of neutron leakage spectra from a graphite cube with a $^{252}$Cf (s.f) source in its centre.

The spatial profile of neutron flux in the central cavity in large graphite insertion was determined using reaction rates. It was reported good agreement between the calculated and measured profile of reaction rates of reactions sensitive in various energy regions in non-boundary positions. Notable discrepancies were observed in boundary regions. At the bottom, they can be attributed to problems in the description of the water-iron boundary. Significant uncertainties in the range of 20% were observed in upper regions. This effect can be attributed to problems with the description of the fission density above the moderator level that have been reported in the past. However, the good description in the central position confirms the applicability of this region for testing activation cross-sections.

**ACKNOWLEDGEMENTS**

Presented results were obtained with the support of these projects: LM2018120 – Reactors LVR-15 and LR-0, TK01030103 – Ex-core Neutron Flux Measurement for the 4$^{th}$. Generation Nuclear Reactors and SANDA project funded under H2020-EURATOM-1.1 contract 847552.